# DISPERSION EQUATION FOR H MODES OF A STRIP LINE WITH A CIRCULAR CYLINDRICAL SHIELD

I. M. Braver, Kh. L. Garb, and P. Sh. Friedberg

*Abstract*–We formulate the dispersion equation for H modes of a circular waveguide with a perfectly conducting strip of zero thickness placed symmetrically in its diametral plane. The wave number of the lowest mode is calculated with high accuracy for various strip widths. Simple asymptotic formulas are derived for the two extreme cases in which the width of the strip is either small or nearly equal to the diameter of the waveguide.

Consider a strip line formed by a circular waveguide with a perfectly conducting strip of zero thickness placed symmetrically in its diametral plane (Fig. 1a). The radius of the waveguide is taken as unity. Let us investigate the dispersion properties of H modes with the same symmetry properties as the $H_{11}$ mode of an empty circular waveguide polarized in the plane $y = 0$. An electric wall may be placed in the plane $x = 0$ for such modes, and the analysis may be confined to the semi-circular waveguide shown in Fig. 1b.

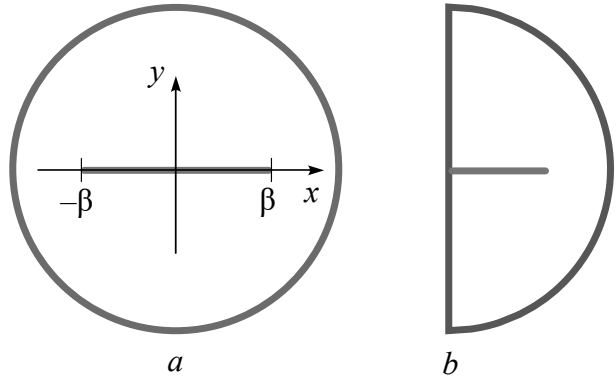

Fig. 1. Cross section of a waveguide with a perfectly conductive strip in the diametral plane. The *z*-axis is perpendicular to the plane of the figure.

The same structure is also obtained upon metallization of the diametral plane of the circular waveguide with internal cruciform conductor that was examined in [1]. The calculations in [1] were performed by a partial-region procedure, without considering the field singularity near the edges of the internal conductor. Therefore, the results of [1] have rather large errors, especially for $\beta \to 1$. The purpose of the present paper is to calculate the dispersion properties of the H modes of a strip line (Fig. 1) for arbitrary strip widths $\beta$.

The necessary dispersion equation can be obtained with the aid of the integral equation (IE) for the density **j** of the surface electric current on the central strip. Since the components of vector **j** are related by $j_z \sim \dfrac{\partial j_x}{\partial x}$ for H modes, it is sufficient to determine only $j_x$. The corresponding IE has the form

$$\int_0^\beta D(x, x_1) j_x(x_1) dx_1 = 0, \quad x \in [0, \beta]. \tag{1}$$

Applying [2, 3], we obtain the following representation for the kernel of this IE:

$$D(x,x_1) = \frac{\kappa^2}{xx_1} \sum_{m,n} \frac{m^2 J_m(\kappa_{hmn}x) J_m(\kappa_{hmn}x_1)}{J_m^2(\kappa_{hmn})(\kappa_{hmn}^2 - m^2)(\kappa_{hmn}^2 - \kappa^2)} -$$

$$- \frac{\partial^2}{\partial x \partial x_1} \sum_{m,n} \frac{J_m(\kappa_{emn}x) J_m(\kappa_{emn}x_1)}{J_{m+1}^2(\kappa_{emn})\kappa_{emn}^2}, \qquad (2)$$

where $\kappa_{emn}, \kappa_{hmn}$ are the positive roots of the equations $J_m(\kappa_{emn}) = 0, J_m'(\kappa_{hmn}) = 0$; $J_m$ and $J_m'$ are the Bessel function and its derivative, and $\kappa$ is the unknown transverse wave number. When the series are summed here and below, the indices $m$ and $n$ assume the values $m = 1, 3, 5, \ldots; n = 1, 2, 3, \ldots$.

We shall seek the solution of the IE (1) in the form of a series

$$j_x = \sum_{v=1}^{N} A_v \psi_v(x) \qquad (3)$$

in a complete system of functions each of which takes account of the known [4] behavior of $j_x$ near the edges of a strip:

$$j_x \sim \sqrt{\beta^2 - x^2}, \quad x \to \pm\beta. \qquad (4)$$

For this system we choose weighted Chebyshev polynomials of the second kind:

$$\psi_v(x) = \frac{4}{\pi}\sqrt{1-u^2} U_{2v-2}(u), \quad u = \frac{x}{\beta}. \qquad (5)$$

Applying the standard procedure of the Galerkin method, we reduce the IE (1) to the dispersion equation

$$\det \|Z_{\mu v}(\kappa)\| = 0, \quad \mu, v = 1, 2, \ldots, N, \qquad (6)$$

$$Z_{\mu v}(\kappa) = \int_0^\beta dx \int_0^\beta dx_1 \psi_\mu(x) D(x,x_1) \psi_v(x_1). \qquad (7)$$

The basic difficulty encountered in subsequent calculations is the need to calculate slowly converging double series. Special methods that accelerate the convergence of such series were developed in [5]. Following [5], we obtain two summation formulas for transformation of the kernel of (2):

$$\sum_{m,n} \frac{J_m(\kappa_{emn}x) J_m(\kappa_{emn}x_1)}{J_{m+1}^2(\kappa_{emn})\kappa_{emn}^2} = \frac{1}{8}\ln\left|\frac{x+x_1}{x-x_1}\right| - \frac{1}{4}\sum_m \frac{(xx_1)^m}{m}, \qquad (8)$$

$$\sum_{m,n} \frac{m^2 J_m(\kappa_{hmn}x) J_m(\kappa_{hmn}x_1)}{J_m^2(\kappa_{hmn})(\kappa_{hmn}^2 - m^2)\kappa_{hmn}^2} = \frac{7xx_1}{32} - \frac{xx_1}{16}\ln|x^2 - x_1^2| +$$

$$+ \frac{1}{8}\sum_m{}' \frac{m^2-2}{m^2-1}(xx_1)^m - \frac{x^2 + x_1^2}{16}\sum_m \frac{m}{m+1}(xx_1)^m, \qquad (9)$$

where symbol ' after the summation sign indicates that the term $m=1$ is excluded. Denoting (8) and (9) by $s^e(x, x_1)$ and $s^h(x, x_1)$, respectively, we represent the kernel of the IE in the form

$$D(x, x_1) = -\frac{\partial^2}{\partial x \partial x_1} s^e(x, x_1) + \frac{\kappa^2}{xx_1} s^h(x, x_1) +$$

$$+ \frac{\kappa^4}{xx_1} \sum_{m,n} \frac{m^2 J_m(\kappa_{hmn} x) J_m(\kappa_{hmn} x_1)}{J_m^2(\kappa_{hmn})(\kappa_{hmn}^2 - m^2)(\kappa_{hmn}^2 - \kappa^2)\kappa_{hmn}^2}. \tag{10}$$

As a result of these operations, the double series in (10) contains the additional multiplier $\kappa_{hmn}^{-2}$ in the general term, which (2) lacks, so that it converges much more rapidly. Evaluation of $Z_{\mu\nu}(\kappa)$ gives rise to the integrals

$$S_{\mu\nu}^e = \int_0^\beta dx \int_0^\beta dx_1 \psi_\mu(x) \left[ \frac{\partial^2}{\partial x \partial x_1} s^e(x, x_1) \right] \psi_\nu(x_1),$$

$$S_{\mu\nu}^h = \int_0^\beta dx \int_0^\beta dx_1 \psi_\mu(x) \frac{s^h(x, x_1)}{xx_1} \psi_\nu(x_1). \tag{11}$$

Recognizing that the logarithmic and power-law functions that appear in (8) and (9) can be integrated analytically [6] with weighted Chebyshev polynomials (5), we find

$$S_{\mu\nu}^e = \frac{2\mu - 1}{2} \delta_{\mu\nu} - (2\mu - 1)(2\nu - 1) \sum_m \frac{\beta^{2m}}{m} \Gamma_{m, 2\mu-1} \Gamma_{m, 2\nu-1},$$

$$S_{\mu\nu}^h = \frac{\beta^2}{32} \left[ \left(7 - 4 \ln \frac{\beta}{2}\right) \delta_{\mu 1} \delta_{\nu 1} + \left(\frac{1}{\mu} + \frac{1 - \delta_{\mu 1}}{\mu - 1}\right) \delta_{\mu\nu} - \frac{\delta_{\mu-1, \nu}}{\nu} - \frac{\delta_{\mu, \nu-1}}{\mu} \right] +$$

$$+ \frac{1}{8} \sum_m{}' \beta^{2m} \frac{m^2 - 2}{m^2 - 1} \left( \Gamma_{m-1, 2\mu-2} - \Gamma_{m-1, 2\mu} \right) \left( \Gamma_{m-1, 2\nu-2} - \Gamma_{m-1, 2\nu} \right) -$$

$$- \frac{\beta^2}{16} \sum_m \beta^{2m} \frac{m}{m+1} \left[ \left( \Gamma_{m-1, 2\mu-2} - \Gamma_{m-1, 2\mu} \right) \left( \Gamma_{m+1, 2\nu-2} - \Gamma_{m+1, 2\nu} \right) + \right.$$

$$\left. + \left( \Gamma_{m+1, 2\mu-2} - \Gamma_{m+1, 2\mu} \right) \left( \Gamma_{m-1, 2\nu-2} - \Gamma_{m-1, 2\nu} \right) \right], \tag{12}$$

$$\Gamma_{m\nu} = \begin{cases} m! \left[ 2^m \left(\frac{m+\nu}{2}\right)! \left(\frac{m-\nu}{2}\right)! \right]^{-1}, & m \geq \nu, \\ 0, & m < \nu, \end{cases} \tag{13}$$

$\delta_{\mu\nu}$ is the Kronecker delta.

The single series in (12) contain the multiplier $\beta^{2m}$ in the general term and converge very rapidly at $\beta < 1$. Therefore, no difficulty is encountered in calculating $S_{\mu\nu}^e$ and $S_{\mu\nu}^h$. We now use the basis functions to integrate the double series in (10). This gives

$$\sigma_{\mu\nu}(\kappa) = \sum_{m,n} \frac{I_{\mu mn} I_{\nu mn}}{J_m^2(\kappa_{hmn})(\kappa_{hmn}^2 - m^2)(\kappa_{hmn}^2 - \kappa^2)}, \qquad (14)$$

where

$$I_{\nu mn} = J_{\frac{m-2\nu+1}{2}}\left(\frac{\beta}{2}\kappa_{hmn}\right)\left[J_{\frac{m+2\nu-3}{2}}\left(\frac{\beta}{2}\kappa_{hmn}\right) - J_{\frac{m+2\nu+1}{2}}\left(\frac{\beta}{2}\kappa_{hmn}\right)\right] +$$

$$+ J_{\frac{m+2\nu-1}{2}}\left(\frac{\beta}{2}\kappa_{hmn}\right)\left[J_{\frac{m-2\nu+3}{2}}\left(\frac{\beta}{2}\kappa_{hmn}\right) - J_{\frac{m-2\nu-1}{2}}\left(\frac{\beta}{2}\kappa_{hmn}\right)\right]. \qquad (15)$$

To derive (15), we used the relation between Chebyshev polynomials of the first and second kinds

$$U_\nu(u)\sqrt{1-u^2} = \frac{T_\nu(u) - T_{\nu+2}(u)}{2\sqrt{1-u^2}} \qquad (16)$$

and the integral (7.361) from [7]. Numerical calculations showed that the double series (14) converges quite rapidly and can be evaluated with an absolute error no greater than $10^{-6}$ for $m, n \leq 50$. Using the quantities found here, the final expression for the dispersion equation matrix elements assumes the form

$$Z_{\mu\nu}(\kappa) = -S_{\mu\nu}^e + \kappa^2 S_{\mu\nu}^h + \frac{1}{4}\kappa^4 \beta^2 \sigma_{\mu\nu}(\kappa). \qquad (17)$$

The results of solution of the dispersion equation for various $\beta$ are assembled in Table 1. The smallest root was found in each case, and the order $N$ of the determinant (6) was made large enough to ensure stabilization of all of the displayed digits of $\kappa$.

Table 1

| $\beta$ | $\kappa$ | $\beta$ | $\kappa$ | $\beta$ | $\kappa$ |
|---|---|---|---|---|---|
| 0.01 | 1.84099 | 0.35 | 1.61349 | 0.70 | 1.16094 |
| 0.05 | 1.83634 | 0.40 | 1.55264 | 0.75 | 1.09486 |
| 0.10 | 1.82172 | 0.45 | 1.48898 | 0.80 | 1.02726 |
| 0.15 | 1.79726 | 0.50 | 1.42378 | 0.85 | 0.956151 |
| 0.20 | 1.76324 | 0.55 | 1.35800 | 0.90 | 0.877387 |
| 0.25 | 1.72040 | 0.60 | 1.29218 | 0.95 | 0.779143 |
| 0.30 | 1.66997 | 0.65 | 1.22654 | 0.98 | 0.689328 |

It was found that when $\beta \leq 0.2$, even the first order ($N = 1$) makes it possible to calculate $\kappa$ with the accuracy of six significant digits. This is because the first basis function $\sqrt{1-u^2}$ becomes the exact solution of the IE (1) as $\beta \to 0$. This fact makes it possible to obtain an asymptotic expression for $\kappa$ that holds for $\beta \ll 1$. Setting $\kappa^2 = \kappa_{h11}^2 - a\beta^2$, where $a$ is a sought for correction factor, we express $Z_{\mu\nu}$ in the form

$$Z_{\mu\nu} = -\frac{1}{2}(2\mu-1)\delta_{\mu\nu} + \frac{1}{a}\frac{\kappa_{h11}^4}{4J_1^2(\kappa_{h11})(\kappa_{h11}^2-1)}\delta_{1\mu}\delta_{1\nu} + o(\beta). \qquad (18)$$

Neglecting $o(\beta)$ terms in (18), we find the correction factor $a$ from (6) and use it to determine $\kappa$ accurate up to terms of orders up to $\beta^2$ inclusive:

$$\kappa = \kappa_{h11} - \frac{\kappa_{h11}^3}{4J_1^2(\kappa_{h11})(\kappa_{h11}^2-1)}\beta^2, \quad \beta \ll 1. \qquad (19)$$

The order $N$ of the determinant must be raised as $\beta$ increases to attain the accuracy of the data in Table 1. The factors that lower the convergence speed of the method with increasing $N$ are described in [8].

Table 2

| $N$ | $\kappa_N$ |
|---|---|
| 1 | 0.705507 |
| 2 | 0.692415 |
| 3 | 0.689896 |
| 4 | 0.689445 |
| 5 | 0.689354 |
| 6 | 0.689334 |
| 7 | 0.689329 |
| 8 | 0.689328 |
| 9 | 0.689328 |
| (27) | 0.6887 |

The data in Table 2 illustrate how the approximations $\kappa_N$ of the unknown wave number $\kappa$ vary as $N$ increases for a broad ($\beta = 0.98$) strip. If $\beta > 0.98$, it becomes difficult to solve dispersion equation (6). But at these values of the parameter $\beta$ it is possible to find $\kappa$ accurately enough from an asymptotic formula derived by solving the IE for the tangential electric field E on the slot between the strip and the surface of the waveguide [9]:

$$\int_\beta^1 G(x, x_1) E_x(x_1) dx_1 = 0, \quad x \in [\beta, 1]. \qquad (20)$$

Here [2]

$$G(x, x_1) = \sum_{k=0}^\infty \varepsilon_k \, g_{2k}(x, x_1),$$

$$g_k(x, x_1) = \frac{Y_k'(\kappa)}{J_k'(\kappa)} J_k(\kappa x) J_k(\kappa x_1) - J_k(\kappa x_<) Y_k(\kappa x_>), \qquad (21)$$

$\varepsilon_k = 2 - \delta_{0k}$, $x_< = \min\{x, x_1\}$, $x_> = \max\{x, x_1\}$, while $Y_k$ and $Y_k'$ are the Neumann function and its derivative. Applying the asymptotic form

$$g_k(x, x_1) \simeq \frac{(xx_1)^k}{\pi k} + \frac{1}{\pi k}\left(\frac{x_<}{x_>}\right)^k, \quad k \to \infty, \tag{22}$$

we represent the kernel $G(x, x_1)$ in the form

$$G(x, x_1) = g_0(x, x_1) + \sum_{k=1}^{\infty}\left\{2g_{2k}(x, x_1) - \frac{1}{\pi k}\left[(xx_1)^{2k} + \left(\frac{x_<}{x_>}\right)^{2k}\right]\right\} -$$

$$- \frac{1}{\pi}\ln\left[1 - (xx_1)^2\right] - \frac{1}{\pi}\ln\left[1 - \left(\frac{x_<}{x_>}\right)^2\right]. \tag{23}$$

We introduce the new variable $v \in [0,1]$ with the relation $x = 1 - (1-\beta)v$ and expand (23) in series in the small parameter $\delta = 1 - \beta$. With accuracy up to terms that do not vanish as $\delta \to 0$ we obtain

$$G(x, x_1) \simeq -\frac{2J_0(\kappa)}{\pi \kappa J_1(\kappa)} + \frac{2\kappa}{\pi}\sum_{k=1}^{\infty}\frac{1}{k}\left[2k\frac{J_{2k}(\kappa)}{J_{2k+1}(\kappa)} - \kappa\right]^{-1} -$$

$$- \frac{1}{\pi}\ln\left(4\delta^2 |v^2 - v_1^2|\right). \tag{24}$$

The function $\dfrac{1}{\sqrt{1-v^2}}$ is the exact solution of IE (20) with kernel (24). Substitution of this function into the IE yields the dispersion equation

$$\ln\frac{1}{\delta} - \frac{J_0(\kappa)}{\kappa J_1(\kappa)} + \kappa \sum_{k=1}^{\infty}\frac{1}{k}\left[2k\frac{J_{2k}(\kappa)}{J_{2k+1}(\kappa)} - \kappa\right]^{-1} = 0, \quad \delta \ll 1. \tag{25}$$

Since $\kappa \to 0$ as $\delta \to 0$, we expand the Bessel functions in (25) in series in the parameter $\kappa$. Accurate up to terms of order $\kappa^2$ inclusive, the series is

$$\ln\frac{1}{\delta} - \frac{2}{\kappa^2} + \frac{1}{4} + A\kappa^2 = 0, \quad A = \frac{\pi^2}{24} + \ln 2 - \frac{95}{96}, \tag{26}$$

from which

$$\kappa = \frac{4}{\sqrt{1 + 4\ln\frac{1}{\delta} + \sqrt{\left(1 + 4\ln\frac{1}{\delta}\right)^2 + 128A}}}, \quad \delta \ll 1. \tag{27}$$

The known expression $\kappa \simeq \sqrt{\dfrac{2}{\ln\dfrac{1}{\delta}}}$, which applies for logarithmically narrow $\left(\ln\dfrac{1}{\delta} \gg 1\right)$ slots, is one implication of (27). The value of $\kappa$ calculated from formula (27) with $\delta = 0.02$ appears in the last line of Table 2. Results calculated using (27) are also plotted in Fig. 2. For comparison, the figure displays the curve constructed using the asymptotic formula (19) and the curve based on the data in Table 1.

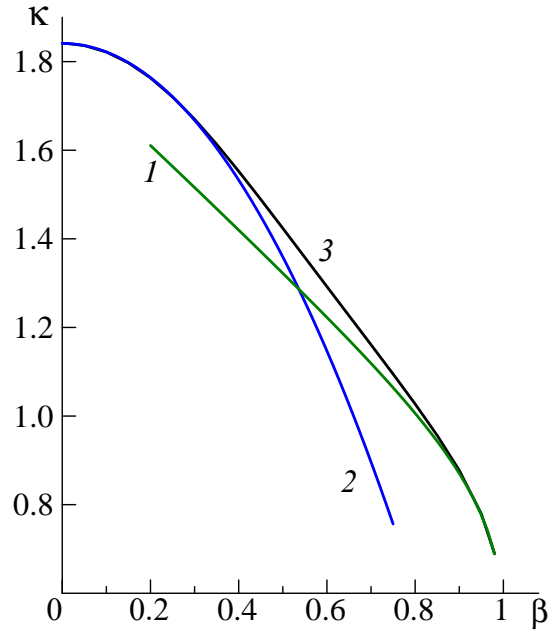

Fig. 2. Transverse wave number of the dominant H mode vs. width of the strip: *1* and *2* – calculations using asymptotic formulas (27) and (19), *3* – solution of the rigorous dispersion equation (6).

We note in conclusion that the strip line considered above was calculated in [10] as the limiting case of a waveguide with a resistive film. Since the basis functions in [10] did not describe the true distribution of current at the edges of the strip (4), the results of that study are less accurate than the data in our Table 1 above. However, comparison indicates that the earlier results have errors no larger than 0.5% for all of the values $\beta < 0.9$ that were considered in [10].

The authors thank Rachel Meyerova for performing the computer calculations for this work.